# Usability and Security of Different Authentication Methods for an Electronic Health Records System

Saptarshi Purkayastha[1], Shreya Goyal[1], Bolu Oluwalade[1], Tyler Phillips[2], Huanmei Wu[3], Xukai Zou[2]
[1]*Department of BioHealth Informatics, Indiana University-Purdue University Indianapolis, Indiana 46202, USA*
[2]*Department of Computer Science, Indiana University-Purdue University Indianapolis, Indiana 46202, USA*
[3]*Department of Health Services Administration and Policy, Temple University College of Public Health, Pennsylvania 19122, USA*
{saptpurk, boluwala, xzou}@iupui.edu, {shregoya, phillity}@iu.edu, huanmei.wu@temple.edu

Abstract: We conducted a survey of 67 graduate students enrolled in the Privacy and Security in Healthcare course at Indiana University Purdue University Indianapolis. This was done to measure user preference and their understanding of usability and security of three different Electronic Health Records authentication methods: single authentication method (username and password), Single sign-on with Central Authentication Service (CAS) authentication method, and a bio-capsule facial authentication method. This research aims to explore the relationship between security and usability, and measure the effect of perceived security on usability in these three aforementioned authentication methods. We developed a formative-formative Partial Least Square Structural Equation Modeling (PLS-SEM) model to measure the relationship between the latent variables of Usability, and Security. The measurement model was developed using five observed variables (measures) - Efficiency and Effectiveness, Satisfaction, Preference, Concerns, and Confidence. The results obtained highlight the importance and impact of these measures on the latent variables and the relationship among the latent variables. From the PLS-SEM analysis, it was found that security has a positive impact on usability for Single sign-on and bio-capsule facial authentication methods. We conclude that the facial authentication method was the most secure and usable among the three authentication methods. Further, descriptive analysis was done to draw out the interesting findings from the survey regarding the observed variables.

## 1 INTRODUCTION

Privacy and security are the major challenges for healthcare organizations to prevent the growing number of data breaches. User authentication is essential in Electronic Health Records (EHRs) to protect sensitive patient data and prevent malicious unauthorized users from gaining access (Jayabalan and O'Daniel, 2019). EHRs include a range of sensitive information, including patient demographics, medical history, immunization status, lab results, radiology images, medications, and identity information like social security numbers, and credit card information. The number of data breaches is rapidly increasing, incurring losses to the health systems worth billions of dollars and identity theft of patients (Ronquillo et al., 2018).

Our study is to shed light on the comparison of the user's perception of different authentication methods used in Electronic Health Records. Authentication can be achieved through various methods. The username-password is the predominant authentication approach used in the information security world (Jayabalan and O'Daniel, 2019). Single sign-on authentication methods refer to a centralized system that allows authenticating to multiple services, without having to remember multiple passwords, reducing password fatigue. Although not very common in healthcare systems, this is starting to be implemented in a few places (Purkayastha et al., 2017). Biometric authentication is also quickly growing in acceptance and used in a wide range of domains. In biometrics, the user needs to be identified based on some characteristic physiological parameters. The purpose is to ensure that the rendered services are accessed only by a legitimate user and not anyone else (Bhattacharyya et al., 2009).

Only a few published studies have focused on developing a proper evaluation framework for the authentication methods' usability and security. Most studies have focused on only one authentication method, like two-factor authentication or biometric authentication (Mihajlov et al., 2011b). In this paper, we built a framework for understanding the interplay of usability and security by using three authentication methods, which provides more generalizability to our framework. Understanding people's attitudes toward security technology is the key to designing both usable and secure systems (Faklaris et al., 2019). We use a mixed-structure survey composed of Likert-scaled questions and open-ended questions to capture user perceptions and build a framework using a robust, statistical approach.

Users perceive that systems designed to increase security are usually less convenient (Ogbanufe and Kim, 2018). Asking users to change their passwords frequently, say every month, reduces the chances of security attacks, but users perceive this to be inconvenient. Therefore, it is essential to study the user perception towards the usefulness, ease of use, and security of the system. Considering these factors to be important, our study investigated the issue to answer the following research questions: How do authentication methods in Electronic Health Records influence: (1) individuals' perceptions in terms of convenience, usefulness, and security concern, and (2) outcomes in terms of the individual's trust and their willingness or satisfaction is continuing to use the authentication method (Ogbanufe and Kim, 2018).

The Biometric authentication method used in the study is the BioCapsule-based role-wise authentication-authorization scheme (Phillips et al., 2017). Our framework might be useful to researchers who wish to improve the usability of their authentication methods. In our study, usability and security are the latent variables. Latent variables (from Latin: present participle of lateo ("lie hidden")), unlike observable variables, are not directly observed but are rather inferred from other directly observed and measured variables through a mathematical model (Cepeda-Carrion et al., 2019). Regarding this, numerous methods and data analysis techniques have been developed to address these issues using unobserved variables (Cepeda-Carrion et al., 2019). One of them is partial least squares structural equation modeling (PLS-SEM), a type of structural equation modeling defined as the combination of latent variables and structural relationships. They are a means by which theoretical relationships can be tested among the variables. The goal of PLS-SEM is to predict the latent (hidden) variables through the values of the observed variables. Our study's observed variables are Efficiency and Effectiveness, Satisfaction, Preference, Concerns, and Confidence.

## 2 RELATED WORKS

While most prior research measures a single authentication system's usability, our work studies perceptions of individuals by comparing three different authentication methods. Over the last few decades, discussion on biometric authentications is common in the information systems and security literature. A literature review by Jayabalan highlights different authentication methods employed in EHRs to protect patient's privacy and security (Jayabalan and O'Daniel, 2019). The study talks about the increase in biometric authentication systems in healthcare organizations as these methods grant access to sensitive Personal Health Information (PHI) only after validating a subject's unique characteristics.

Another study presents a survey to measure the user perception of Duo Two-Factor-Authentication (2FA) at Brigham Young University (Dutson et al., 2019). The survey questions gathered both qualitative and quantitative data. Another study (Hewitt and McLeod, 2011) performed a survey to better understand the impact of the implementation of security features like biometric authentication on the ease of use and usability. The study results highlight that the benefit of quick access to the patient data via biometric authentication influenced their opinions about using an EHR system, speeding up the access process. The study also addressed some of the challenges while implementing biometric authentication. These include providers wearing gloves, masks, and goggles that may hinder the biometric authentication method. Also, dirt, grime, grease, blood, and cleaning solutions can reduce biometric scanner reliability. Another research has presented a quantification approach for assessing usable security in authentication mechanisms with the purpose of guiding the usability and security evaluation process. The proposed framework focuses on the system's quality in terms of its suitability for the task in a particular domain (Mihajlov et al., 2011a).

Most studies that include biometric authentication are conceptual in nature and lack theoretical foundations. We did not find much research comparing biometric authentication with other authentication methods in the Electronic Health

Records systems, particularly for usability, usefulness, and security. To the best of our knowledge, this paper is the first study to compare user perception of biometric authentication with other authentication methods to explain the interplay between usability and security.

## 3 METHODOLOGY

We conducted an IRB-approved study #2002481071. The survey contained 30 questions designed to identify the Effectiveness of the authentication methods, User satisfaction, Preference towards the authentication methods, Common usability and Security concerns.

We deployed the popular, open-source electronic health records system, OpenMRS, on three Virtual Machine (VM) instances on the JetStream infrastructure. Each VM had one authentication method - a single-factor authentication (password) method, a single sign-on authentication method with the Indiana University Central Authentication Service (CAS), which is a 2FA method, and a bio-capsule facial authentication method for logging into the OpenMRS system with demo patient information.

We created accounts for all the students enrolled in the Privacy and Security in Healthcare course in all the 3 OpenMRS instances. The credentials were provided to them individually. The students were asked to register themselves using their face and the username provided for the bio-capsule face authentication instance. This registration step is required only once. After the registration is complete, the students can simply authenticate themselves by making their face available in front of the camera. We provided detailed instructions to the students for the registration process. Figure 1 describes the experimental design. The students can access the system as many times they wish to using the credentials provided.

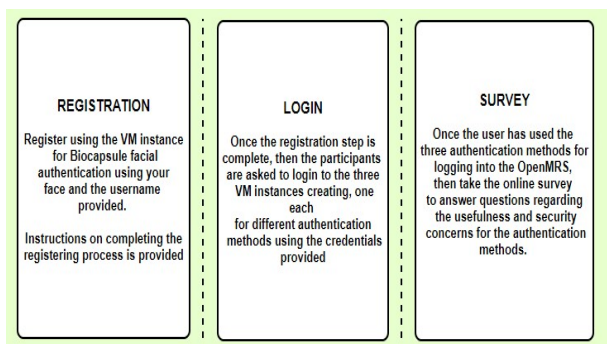

Figure 1: Experiment Design

### 3.1 Data Collection

Qualtrics was used to create the online questionnaire and subsequently collect the results from participants. The use of the Qualtrics allowed for reduced chances of input errors as well as increased access to participants by the researcher (Ogbanufe and Kim, 2018). The students authenticated themselves and logged in to each instance using the credentials provided. Based on their experiences with the authentication process, the students were asked to complete the survey. The survey contains 5-point Likert scale questions, ranking questions, and qualitative questions regarding the efficiency and effectiveness, satisfaction, preference, concerns, and confidence of the three authentication methods studied. The survey was accessible online through a link provided to the students along with the EHR credentials. The survey was open for a period of 10 days.

The first few questions collected demographic information and technical capabilities, including information about past experience with EHR systems, prior experience with authentication methods, and areas of expertise. All the participants were 18 years or above. Most of the participants belongs to the 25-34 age group. Out of the total 33 participants selected for the study, 19 were female and 14 were males.

The professional background/experience of the respondents in percentages was as follows:

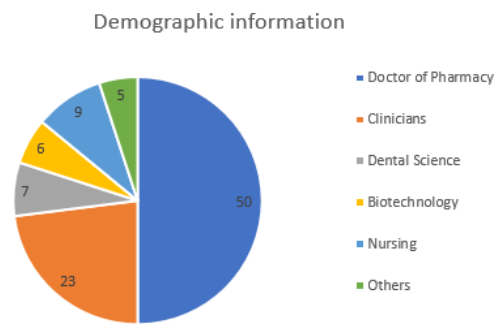

Figure 2: Demographic information

The others include students from Computer science and other engineering and technology-related backgrounds. Figure 3 describes the information regarding participants' prior experience with the EHR authentication methods.

Most participants had prior experience with the Username/password authentication method, followed by the single sign-on and least prior experience with facial authentication method.

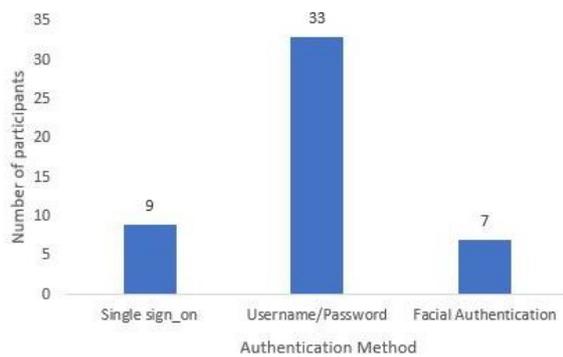

Figure 3: Prior Experience with EHR authentication methods

We divided the survey into sections containing questions about the efficiency and effectiveness, satisfaction, preference, concerns, and confidence of each authentication method. We included three qualitative questions to dig deeper into the responses regarding authentication failure, challenges faced, and authentication method preferences.

## 3.2 Research Model and Hypothesis Development

We used Partial Least Squares Structural Equation Modeling (PLS-SEM) for statistical analysis. Based on the survey responses, we built three different formative-formative type PLS-SEM models - one for each authentication method, using the SmartPLS 3 software. These models explored the affect of perceived security on usability (structural model) and understand the relationship between the observed variables on the latent variables (measurement model). Figure 4 outlines the structural equation model used in the study. Latent variables are drawn as circles, whereas the observed variables determined from our survey questions (items) are shown as rectangles. The detailed description of all the items and its associated observed and latent variables is shown in Table 1. The questions that were not included in the model include the ranking and qualitative questions. Also, we removed highly correlated/similar quantitative questions, which had no significant impact on the results, through an iterative process of model building and testing for convergence.

Our hypothesis were based on the impact of the observed variables for different authentication methods and meant to understand the relationship between usability and security. We hypothesize:

- H1: Efficiency & Effectiveness is believed as a

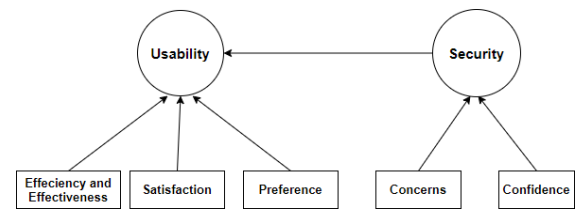

Figure 4: Structural Equation Modeling

Table 1: Cognitive component of the Partial Least Squares model

| Latent variable | Observed variable | Items |
|---|---|---|
| Usability | Efficiency and Effectiveness | Q4. Does not add significant additional time<br>Q5. Easy configuration<br>Q6. User interface: Always know what to do<br>Q7. Efficient<br>Q9. Easy to use<br>Q10. No authentication failure |
| | Satisfaction | Q11. Convenient<br>Q14. Enjoyed<br>Q16. Effortless |
| | Preference | Q19. Willingness to continue using it |
| Sceurity | Concerns | Q24. EHR account not being compromised when using this method<br>Q25. No unauthorized personnel can gain access |
| | Confidence | Q26. Feel secure<br>Q27. Comfortable using the method as primary EHR authentication method<br>Q28. Not have to worry about the account safety |

serious concern/important factor in determining the usability of authentication methods.

- H2: High Satisfaction is believed as a serious concern/important factor in determining the usability of authentication methods.

- H3: Security concerns is believed as an important factor in determining the security of authentication methods.

- H4: confidence/willingness to continue using is believed as a serious concern/important factor in the security of authentication methods.

- H5: Security have a significant impact on the usability of the authentication methods for Electronic Health Records.

We used the repeated indicator approach where a higher-order latent variable is constructed by specifying a latent variable that represents all the manifest variables of the underlying lower-order latent variables (Becker et al., 2012). We measured the observed variables by asking respondents a 5 point Likert questions related to the construct represented as the items with responses ranging from 1 to 5 with 1 indicating strongly agree and 5 indicating strongly disagree with the question statement.

# 4 DATA ANALYSIS AND RESULTS

Overall, we received responses from 57 participants, achieving a response rate of about 84%. After removing the incomplete responses, we were left with 37 responses (i.e approx. 55% response rate). This section will present the insights from the model, qualitative and quantitative analysis, and a discussion of our results in the context of usability and security.

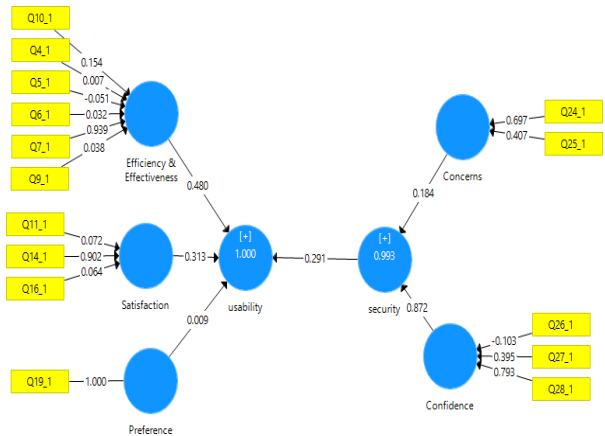

Figure 5: Facial authentication model

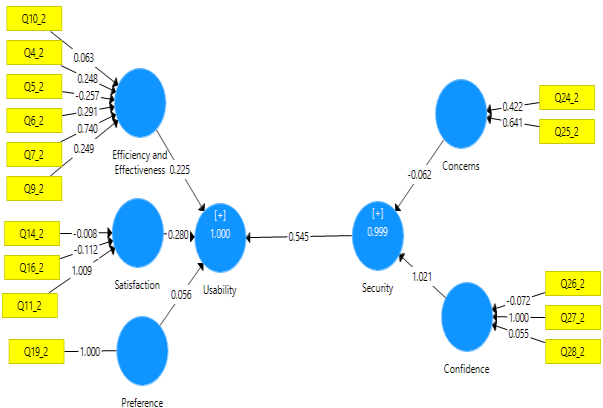

Figure 6: Single sign-on model

Figures 5, 6, and 7 describe the three PLS-SEM models with the path coefficient for the three authentication methods. The path coefficient in PLS-SEM is interpreted as a standardized regression coefficient: If $X$ changes by one standard deviation, then $Y$ changes by $b$ standard deviations (with $b$ being the path coefficient).

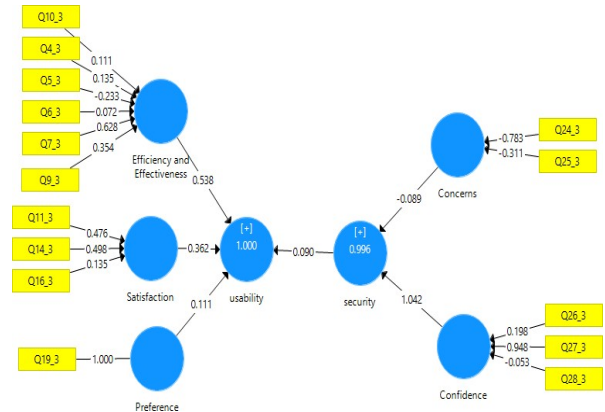

Figure 7: Username/Password model

## 4.1 Quantitative Analysis

Even though we could have studied the effect of usability on security, it made less intuitive sense to us and instead we studied the impact of security on usability for the three different authentication methods.

### 4.1.1 PLS-SEM Analysis

**1. Assessment of First-Order Construct Model:** In order to assess the validity of formative constructs at the first-order construct level, several steps need to be done sequentially. The first step is to assess the multicollinearity issue in the formative indicators. High correlations between two formative indicators can have an effect on the results. In the context of PLS-SEM, a variance inflation factor (VIF) value of 5 and higher indicates a potential collinearity problem (Hair et al., 2012). Five formative constructs (efficiency and effectiveness, satisfaction, preference, concerns, and confidence) were assessed for validity using the VIF values. All the VIF values were less than 3; hence the first assessment step for the formative constructs was met.

The item weight in the formative measurement models was analyzed for their significance and relevance as the second step for formative measurement assessment (Hair et al., 2012). We used a bootstrapping procedure generating 1000 subsamples to test whether the outer weights in formative measurement models are significantly different from zero. These weights are described in figures 5, 6, and 7. According to Diamantopoulos and Winklhofer (2001), even an insignificant indicator should be preserved in the item set capturing the construct since it may still represent some of the domain aspects (Andreev et al., 2009). Hence, all the items were preserved for the second-order construct analysis.

**2. Assessment of Second-Order Construct Model:** In order to assess the formative second-order construct, the collinearity issue was addressed as mentioned earlier. Next, we assess the formative indicators for the second-order construct model. Tables 2 and 3 describe the path coefficients and T-statistics for the second-order constructs. We observed that among the five indicators, the significant ones were efficiency and effectiveness, satisfaction, and confidence for the facial authentication and username/password methods, while satisfaction and confidence were significant for the single sign-on authentication method. Values above 1.96 are considered to have a significant impact. The model was evaluated by calculating the R-squared value for each of the latent variables. From Figures 5,6 and 7, the R-squared values were found to be close to 1, which defines excellent predictive accuracy.

Table 2: Path coefficients for all the three methods

| Second order construct | First order construct | Path coeff (facial auth) | Path coeff (single sign-on) | Path coeff (user/ pass) |
|---|---|---|---|---|
| Usability | Efficiency & Effectiveness | 0.480 | 0.225 | 0.538 |
|  | Satisfaction | 0.313 | 0.280 | 0.362 |
|  | Preference | 0.009 | 0.056 | 0.011 |
| Security | Concerns | 0.184 | -0.062 | -0.089 |
|  | Confidence | 0.872 | 1.021 | 1.042 |

Table 3: T-statistics for all the three methods

| Second order construct | First order construct | T-stats (facial auth) | T-stats (single sign-on | T-stats user/ pass |
|---|---|---|---|---|
| Usability | Efficiency & effectiveness | 4.933* | 1.839 | 4.896* |
|  | Satisfaction | 2.249* | 2.466* | 3.507* |
|  | Preference | 0.292 | 0.611 | 1.321 |
| Security | Concerns | 0.437 | 0.575 | 0.503 |
|  | Confidence | 4.262* | 16.330* | 8.353* |

Thus, the model explained 100% of the construct's variance for usability and around 99% for security. The impact of security on usability was found to be significant in the case of facial authentication and single sign-on with t-stats values of 2.363 & 3.612 respectively highlighting a positive significant impact of security on usability.

From the Table 4, we can conclude that the H1 is partially accepted, in the case of facial authentication and username/password authentication methods. Efficiency & effectiveness are considered to have a significant impact on the usability aspect.

Table 4: Assessment of structural model for hypothesis testing

| Hypothesis | t-value (face_auth) | t-value (single sign-on) | t-value (user/ pass) | Supported? |
|---|---|---|---|---|
| H1 | 4.933* | 1.839 | 4.89* | Partial |
| H2 | 2.25* | 2.47* | 3.5* | Yes |
| H3 | 0.43 | 0.57 | 0.5 | No |
| H4 | 4.26* | 16.33* | 8.35* | Yes |
| H5 | 2.36* | 3.61* | 1.86 | Partial |

H2 is accepted that satisfaction has a significant impact on the usability of all the three authentication methods. H3 is rejected as it was found that the security concerns for the authentication methods were not found to significantly impact the security aspect. The H4 is supported as the confidence/willingness to continue using the same method significantly impacted security. The last hypothesis, which states that security has a significant impact on usability, is partially supported, at least in the case of facial authentication and single sign-on authentication methods.

From the path coefficient values, it can be observed that the security concerns are seen to have a positive association with the security for facial authentication, indicating that the users feel more secure using facial authentication methods when compared to other authentication methods. The security concerns for the other two methods are seen to be negative, indicating that the users are not feeling secure while using those authentication methods. The other interesting observation was the magnitude of the path coefficient for confidence. It can be observed that the confidence among the users for continuing to use the method was significantly high in the case of facial authentication when compared to others. This shows that the users are willing to adapt and continue using the facial authentication methods for logging into Electronic Health Records. This observation supports biometrics literature that suggests that biometrics is more secure. The results from the model justify our findings from the survey, where the respondents found the facial authentication and single-sign-on method to be more secure and usable when compared to the username/password authentication method.

## 4.2 Descriptive Analysis

Usability is measured by three factors - Efficiency and Effectiveness, Satisfaction, and Preference. Figures 8 and 9 depict the response percentage in the order of their ranking, with 1 being the least efficient and preferred and 3 being the most efficient and preferred authentication method. Facial authentication was found to be the most efficient as well as the most preferred authentication method followed by the single sign-on authentication method.

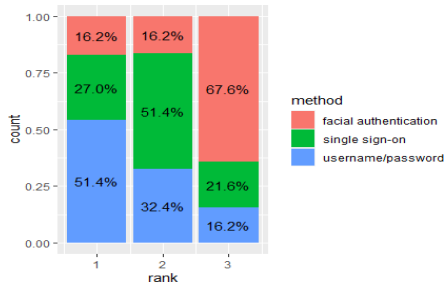

Figure 8: Comparing Efficiency and Effectiveness among authentication methods

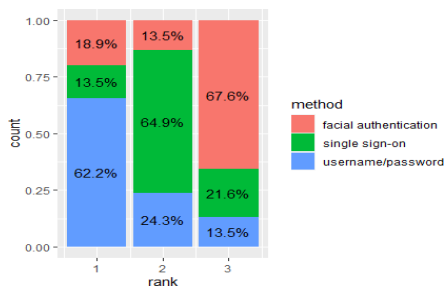

Figure 9: Comparing Preference among authentication methods

Though the facial authentication was found to be the most efficient and preferred authentication method, respondents found the configuration process to be cumbersome. It was observed that around 60% of the respondents agreed that they found the configuration method of facial authentication to be cumbersome. But once configured, they found it to be quick, efficient, and easy to use.

As regards security concerns, more than 90% of the respondents were confident using the facial authentication method and believed that this method was secure against any security attacks when compared to the other two methods. The least confident method in terms of security was found to be the username/password method with 41.2% of the respondents not feeling secured while using it.

### 4.3 Qualitative Analysis

This section describes in detail the challenges faced by the respondents while using the authentication method, the most and the least liked feature in the authentication methods. Some of the interesting comments that were received:

*"Face authentication is fast and easy because I don't have to type or say anything."*
*"My face is only my face. Works hands-free."*
*"Face-Authentication is my most preferred authentication method. It is easy to use, enhances the security, and avoids fraudsters hacking the EHR system using this method."*
*"Face authentication is fast and easy because do not have to type or say anything."*

These comments describes the features liked about facial authentication is its ease of use as there is no need to remember complex passwords and secure as there are fewer chances of hacking and other security breaches.

We also asked them to mention the feature they disliked in the authentication method. Some of the comments are: *"Key-loggers can capture password typing and more chances of a data breach in username/password method"*
*"Can forget a difficult password or have to copy-paste from message or file"*
*"Complex password is hard to recall"*
*"Same passwords are commonly used between website, more chances of the data breach"*

Multiple respondents expressed disliking the username/password authentication method for the need to remember complex passwords and increased chances of sensitive health data breaches. However, some users highlighted issues with face authentication. The following comment highlights that camera is an essential requirement for face authentication that might disenfranchise certain users - *"I opted for face authentication as least preferred because I was not able to use the face authentication as my laptop does not have a camera facility. So maybe it would be difficult for everyone to have access to this method"*.

The other response question recorded the challenges faced by the respondent while authenticating themselves. The challenges faced are described below:

- I faced difficulty while using face authentication as my laptop do not have a camera.

- My camera was set up for the rear-facing camera, so I could not easily get a photo of myself. I did not see any options to change camera settings in the program.

- Going back again and again to get the complex password.

The challenges described by the respondents is quite interesting, as it gives credence to the use of non-facial authentication methods because its use is not widespread at the moment, while the username/password and single sign-on method are quite ubiquitous.

## 5 CONCLUSION

In this study, we compared three different authentication methods deployed in the OpenMRS EHR in terms of usability and security. We also examined how security features such as biometrics and single sign-on impact the user acceptance of the authentication method. Compared to the other studies (Hewitt and McLeod, 2011), our study also highlights the positive impact of security features on usability and acceptance of the authentication method for Electronic Health Records. Analysis of the survey results shows the facial authentication method was perceived as the most efficient and preferred authentication method. Our model showed that there is a significant positive relationship between perceived efficiency and preference.

The findings presented in our model shows that for the facial authentication and single sign-on methods, there is a significant positive impact of security on usability. This impact was not found in the username/password authentication method. Based on our result, we conclude that the facial authentication method was the most secure and usable. This finding is unsurprising given that one of the major advantages of the biometric authentication method is its combination of convenience of use and difficulty in forging an access pattern.

## ACKNOWLEDGEMENTS

The U.S. National Science Foundation supported this work under grant OAC-1839746. This work was made possible through a research allocation on the JetStream (Stewart et al., 2015) public cloud infrastructure and XSEDE resources.